\newcounter{myctr}

\documentclass{ws-acs}
\usepackage{url}
\usepackage{amssymb}
\usepackage{setspace}
\usepackage{multirow} 
\usepackage[table]{xcolor}
\begin{document}

\makeatletter
\def\@biblabel#1{[#1]}
\makeatother

\markboth{Authors' Names}{Instructions for
Typing Manuscripts (Paper's Title)}

%
\catchline{}{}{}{}{}
%

\title{Complex Network Analysis in Cricket : Community structure, player's role and performance index}

\author{\footnotesize Satyam Mukherjee}

\address{Kellogg School of Management, Northwestern University\\
Evanston, IL
United States of America\\
satyam.mukherjee@gmail.com}

\maketitle


\begin{abstract}
This paper describes the applications of network methods for understanding interaction within members of sport teams.We analyze the interaction of batsmen in International Cricket matches. We generate batting partnership network (BPN) for different teams and determine the exact values of clustering coefficient, average degree, average shortest path length of the networks and compare them with the  Erd\text{\"{o}}s-R\text{\'{e}}nyi model. We observe that the networks display small-world behavior. We find that most connected batsman is not necessarily the most central and most central players are not necessarily the one with high batting averages. We study the community structure of the BPNs and identify each player's role based on inter-community and intra-community links. We observe that {\it Sir DG Bradman}, regarded as the best batsman in Cricket history does not occupy the central position in the network $-$ the so-called connector hub. We extend our analysis to  quantify the performance, relative importance and effect of removing a player from the team, based on different centrality scores. 
\end{abstract}

\keywords{Complex network; Small world behavior; Centrality scores; Cricket.}


\section*{Introduction}

In recent years there has been an increase in study of activities involving team sports. Time series analysis have been applied to football \cite{naim05,bittner}, baseball \cite{petersen08,sire09}, basketball \cite{naim07,skinner10,guerra} and soccer \cite{Rubner,Ribeiro}. Again, a model-free approach was developed to extract the outcome of a soccer match\cite{heuer10}. The study of complex networks have attracted a lot of research interests in the recent years \cite{laszlo,freeman}. A salient feature of such complex networks is that they display {\it small-world behavior} \cite{watts}. 
The tools of complex network analysis have previously been applied to sports. Such as a network approach was developed to quantify the performance of individual players in soccer \cite{duch10} and football \cite{pena}. Network analysis tools have been applied to football \cite{girvan02}, Brazilian soccer players \cite{onody04}, Asian Go players \cite{xinping}. Successful and un-successful performance in water polo have been quantified using a network-based approach \cite{mendes2011}. Head-to-head matchups between Major League Baseball pitchers and batters was studied as a bipartite network \cite{saavedra09}. More recently a network-based approach was developed to rank US college football teams \cite{newman2005}, tennis players \cite{radicchi11} and cricket teams and captains \cite{mukherjee2012}. 

The complex features of numerous social systems are embedded in the inherent connectivity among system components \cite{laszlo,mendes2011}. Social network analysis (SNA) allow researchers to explore the intra-group and inter-group relations between players, thus providing an informal relation between various players. Such an analysis provides insight about the pattern of interaction among players and how it affects the success of a team. This article points out that topological relations between players need to be explored in order to better understand individuals who play for their teams. SNA is well suited to investigate the complex relations between team members \cite{Lusher}. Such an approach to cricket at the microscopic level, form a basis of elucidating the individual importance and impact of a player. 

Cricket is a game played in most of the Commonwealth countries. In this paper we consider top six Test teams in the history of Test cricket \footnote{There are three versions of the game - `Test', One Day International (ODI) and Twenty20 (T20) formats. Test cricket is the longest format of the game dating back to $1877$. Usually it lasts for five days involving $30-35$ hours. Shorter formats, lasting almost $8$ hours like ODI started in $1971$ and during late $2000$ ICC introduced the shortest format called T20 cricket which lasts approximately $3$ hours.} - Australia, England, South Africa, West Indies, Pakistan and India \cite{mukherjee2012}. 
In the recent years there has been a growth in research interests on Cricket. Multiple linear regression techniques were used to determine relative batting and bowling strengths in  ODI and Test cricket \cite{allsopp}. Again, the distribution of runs scored in Test cricket followed a negative binomial distribution \cite{philip}. Recently it was shown that the score dynamics in cricket is an anomalous diffusive process \cite{RibMukHan}. However these studies do not focus on the style of play adopted by different teams. All renowned Cricket teams have different approach to their game-play, batting line-up and bowling order. Teams like Australia, England rely more on fast bowlers while subcontinent teams like India and Sri Lanka depend on slow bowlers. Again batting order of teams like Australia is much different from teams like West Indies or India. In this paper, we apply tools from SNA to understand a Cricket team's style of play, relative importance of each player within a team and effect of removing a player from the team. The rest of the paper is presented as follows : In Section 2 we propose the methods of link formation among the batsmen. In section 3 we discuss the topological properties and community structure of batsmen partnership network and player's individual role in the team and we conclude in Section 4. 

\section*{Cricket as a complex network}

{\bf The central goal of network analysis is to capture the interactions of individuals within a team. Teams are defined as groups of individuals collaborating with each other with a common goal of winning a game. Within teams, every team member co-ordinate across different roles and the subsequently influence the success of a team. In the game of Cricket, two teams compete with each other. Although the success or failure of a team depends on the combined effort of the team members, the performance or interactive role enacted by individuals in the team is an area of interest for ICC officials and fans alike. We apply network analysis to capture the importance of individuals in the team. The game of Cricket is based on a series of interactions between batsmen when they bat in partnership or when a batsman is facing a bowler. Thus a connected network among batsmen arises from these interactions. }

Cricket is a bat-and-ball game played between two teams of $11$ players each. The complex nature of this sport demarcates it from other sports like soccer or baseball. For example, although baseball appear to be similar to cricket, they share notable differences in the manner in which they are played. In cricket there are many factors which determine the outcome of a game. For example in a cricket line-up, the openers lay the foundation of an innings, by seeing-off the new ball and playing a sheet anchor role. The lower-middle depending on the score either tries of make as many runs as possible or try to save their wickets. From the perspectives of network theory, while baseball pitchers and batters could be investigated as a bipartite network (as in Ref. \cite{saavedra09}), in cricket one cannot represent batsmen and bowlers as a bipartite graph. This is due to the fact that a bowler has to bat once the top order batsmen are dismissed. {\bf Again, sometimes a batsman could be used as a part time bowler.}

 The team batting first tries to score as many runs as possible, while the other team bowls and fields, trying to dismiss the batsmen. At the end of an innings, the teams switch between batting and fielding. In cricket two batsmen always bat in partnership. Usually the opening partnerships are responsible to face the `new ball\rq{} and score runs at the same time. Middle order partnerships are entrusted with consolidation of the innings. Lower order partnerships are much smaller than the opening partnerships. The outcome of a match depends on the batting partnerships between batsmen. Large partnerships not only add runs on the team\rq{}s score, it may also serve to exhaust the tactics of the fielding team. Again, the concept of partnerships become vital if only one recognized batsman remains. It is therefore important to identify the key players in a team by constructing network of batting partners ({\bf See Supporting Information}). We analyze the data of batting partnership (publicly available in cricinfo website \cite{cricinfo}) in Test cricket between $1877$ and August $2012$. Two batsmen are connected if they formed a batting partnership in at least one match. An undirected and unweighted batting partnership network is generated for each country \footnote{Note that partnership networks are always restricted to countries $-$ two batsmen of different countries do not bat together}.

\subsection*{Topological analysis of the network}
\begin{figure*}
\begin{center}
\includegraphics[width=14.0cm]{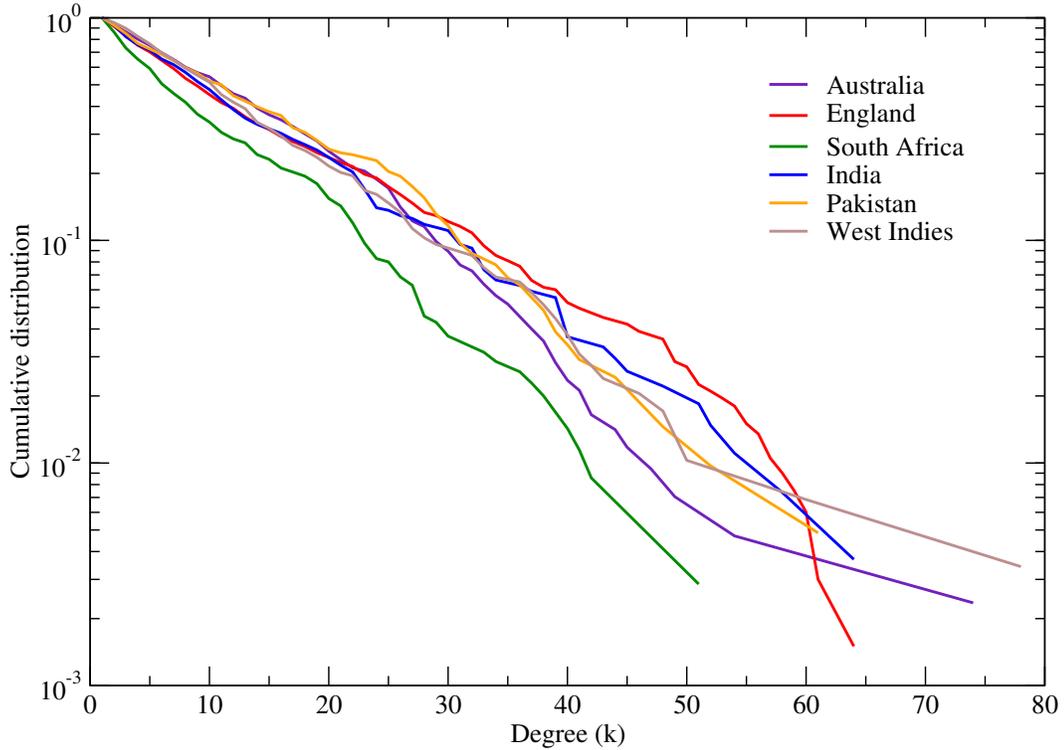}
\caption{ \label{fig:deg2}(Color online) Cumulative degree distribution of batting partnership network for different Teams. }
\end{center}
\end{figure*}

In this section we analyze the batting partnership as a complex network of interaction of two batsmen.  We analyze the topological properties of the interactions among batsmen for different teams in Test cricket ($1877-2012$). We observe that the Australian BPN has $425$ nodes and $2827$ edges.  The average degree $\bar{k}$ is $13.30$. Degree of a node is the number of nodes it is directly linked to. Degree is one of the  centrality measures of a network \cite{wasserman}. The degree distribution reflects the topology of the network and how the batsmen interact with each other during partnerships. In Figure ~\ref{fig:deg2} we plot the degree distributions for different teams. It's evident from the figure that the distribution is neither normal nor a power-law. As we can see, the degree distribution for South Africa decay faster than other teams. This is supported by the fact that between $1971$ and $1991$ South Africa was banned from international Cricket. This prevented inclusion of new players and hence batsmen were not able to form partnership more frequently compared to other teams. We also observe that for England the degree distribution exhibit truncated behavior with a cut-off at $k \sim 50$. The distributions decay slowly for smaller values of $k$, while they decay faster for larger values of $k$. The truncated behavior is justified by the fact that in the initial stages there are fewer individuals who plays the game for England. With increase in popularity more and more matches are being played and the number of batsmen who interact with one another increase. This leads to increase in connections between old and new players. This growth mechanism could account for the nature of the degree distribution. 

We analyze the topological properties of the {\it batsmen partnership network} (BPN). Clustering coefficient ($C_{i}$) of a node $i$ is defined as the ratio of number of links shared by its neighboring nodes to the maximum number of possible links among them. The average clustering coefficient is defined as,  

\begin{equation} 
C = \frac{1}{N}\displaystyle\sum\limits_{i=0}^N C_{i}
\end{equation}
The average clustering coefficient  ($C$) captures the global density of interconnected nodes in a network. For BPN of Australia we observe that $C = 0.60$, indicating that the network is highly clustered.  Keeping the number of nodes fixed, we compare the results with random graphs generated according to the Erd\text{\"{o}}s-R\text{\'{e}}nyi model \cite{erdos}.The clustering coefficient ($C = 0.03$) is seen to be lower than that of the original network.

We evaluate the degree correlations of the BPN. The degree-degree correlation function $A$ (assortativity coefficient) which measures the tendency of a network to connect vertices with the same or different degrees  \cite{newmanA}. Mathematically, its defined as
\begin{equation} 
A = \frac{1}{\sigma_{q}^2}\displaystyle\sum\limits_{jq} jk(e_{jk}-q_{j}q_{k})
\end{equation}
where $q_{k}$ = $\displaystyle\sum_{j}e_{jk}$ and $\sigma_{q}^2$ = $\displaystyle\sum_{k} k^{2}q_{k}$ - ${\displaystyle\sum_{k} kq_{k}}^2$. Here $e_{jk}$ is the probability that a randomly chosen edge has nodes with degree $j$ and $k$ at either end. 
If $A > 0$ ($A < 0$) the network is said to be assortative (disassortative). It has been observed that social networks are assortative and technological and biological networks are disassortative \cite{sen}. For the BPN of Australia  we find that $A = -0.09$, {\bf indicating that weak disassortativity of BPNs.} 

There has been great interest in the shortest path length among nodes in networks \cite{wasserman}, \cite{peay}. We evaluate the average shortest path length $L$ between a given node and all other nodes of the network. For BPN of Australia  we get $L = 5.25$. We observe that the average shortest path length of these networks are of the same order of the corresponding random graphs generated by the ER model ($L=2.59$). Thus the networks display small-world properties. Similar small world phenomenon was earlier observed in World Professional Tennis Players \cite{hokky}. 

We find that the most connected players are not necessarily the most central - players through which most shortest paths go. Also the most `central' players are not necessarily the players with higher batting averages (See Table~\ref{tableS1}). For example, Australia's {\it Sir DG Bradman} holds the highest batting average of $99.94$ in Test cricket. However he is not the player with highest betweenness centrality. For BPN of Australia, {\it RN Harvey} emerge as the most central player. Similarly {\it AR Border} has the highest degree ($k=74$) and yet has a low betweenness centrality of $0.148$. Again {\it A Ward}, {\it JP Duminy} and {\it LR Gibbs} do not boast of a high batting average, but they have high betweenness centrality. On the contrary {\it Javed Miandad} holds the highest batting average for Pakistan and is the most central player. Similar situation is seen for India's {\it P Roy} who was an important batsman during his playing days ($1951-1960$) even though his average is much below than that of modern day Indian batsmen. This also reflect how for subcontinent teams batting performance depends on individual performance of top order batsmen while other teams rely more on team game.

Such a situation of central nodes with low degree was observed earlier in the context of transportation networks \cite{amaralpnas}. In terms of cricket such anomalous centralities arise when different players form groups. We identify communities in the BPN of each team and characterize the role of each player based on its intra-community and inter-community linkages and not merely on the degree or betweenness alone. 

\subsection*{Community structure and roles}
 In the context of social networks, communities \cite{girvan02,newman03,boguna_2004,guimera03,guimera04,reichardt06,newman06b,pons_walktrap_2005,Clauset2004} are groups of individuals who are more densely connected to one another than to the remainder of the network and community detection in networks has been extensively studied over the last few years.  
 {\bf Detecting communities in networks help in identifying functional subunits of the networks. It also helps in uncovering important features in the network which are not apparent in absence of detailed information. 
 Nodes that belong to the same community are classified according to their position within the community. This also gives an idea about the role played by each node. For example nodes present in the core of a cluster plays a role different than those present in the boundary. Again, visualization of networks are almost impossible in many large systems. Community structure provides us with a powerful visual representation of the networks. In terms of sport it is thus important to elucidate the information of players and role they play and community detection tools serve as a handy tool for our study. }
 
 We analyze the community structure of the giant component of the BPN for different teams, via the modularity maximization approach proposed by Newman \cite{newmancom} and we validated these results using different additional methods - random walk algorithm proposed by Rosval et. al. \cite{rosval} and OSLOM (Order Statistics Local Optimization Method) \cite{oslom2} (See Table~S6 and Table~S7 in SI ). We found at least $4$ communities in the giant component (GC) of the BPN for various teams (see Table~\ref{tableS2}). We evaluate the value for modularity $Q$ for each BPN belonging to different teams and compare the actual community structure with that of 100 randomized networks of equal size and degree distribution\cite{maslov02}. We found that the  $z$-score$>2.0$, suggesting that the community structure in the network is statistically significant. In Table~\ref{tableS6} we show the clusters obtained after implementing Rosval's algorithm \cite{rosval}. 
In Table~\ref{tableS2} and Table~\ref{tableS6} we show the number of members of each community, mean batting average, and the 95\% confidence interval for the each community. We observe that the 95\%  confidence intervals for all communities overlap, which suggests that there is no statistically significant difference in the mean batting average between communities. We observe that although both community detection algorithms  give different number of communities one fact stands out -  contemporary players who form batting partnership more often are more connected to one another than other players. {\bf This is again explained by the fact that top-oder batsmen form partnership with each other than more often than with lower-order batsmen. Similarly middle order batsmen partner with top-oder batsmen as well as lower-order batsmen (See SI for the cricketing terms).} For example {\it Sir DG Bradman}, {\it SJ McCabe} and {\it WH Ponsford} played during the same era and belong to the same community. Again, India's {\it SM Gavaskar} and {\it N Kapil Dev} played during same era but belong to different communities. This is due to the fact that both these players never formed batting partnerships too often with each other. Similar arguments hold for West Indies' {\it BC Lara} and {\it S Chanderpaul}.

Having obtained the community structure of our network, we classified the nodes according to the role that they play both within and outside their community. Following the role classification approach presented by Guimer\'a et al.\cite{guimera05} we first made a distinction between hubs and non-hubs. We evaluate the within-community degree $z$-score given by the following equation

\begin{equation}
z_{i} = \frac{\kappa_{i} - \kappa_{s_{i}}}{\sigma_{\kappa_{s_{i}}}}
\end{equation}
where $\kappa_{i}$ is the number of edges of node $i$ to other nodes in its community $s_{i}$, $\kappa_{s_{i}}$ is the is the average of $\kappa$ over all of the nodes in $s_{i}$, and $\sigma_{\kappa_{s_{i}}}$ is the standard deviation of $\kappa$ in $s_{i}$. Next we distinguish nodes based on their connections to nodes belonging to different communities \cite{guimera05}. As defined in Ref.~\cite{guimera05}, the participation coefficient $P_{i}$ of a node $i$ is given as 

\begin{equation}
P_{i} = 1 - \displaystyle\sum\limits_{s=1}^{M} (\frac{\kappa_{is}}{k_{i}})^{2}
\end{equation}
where $\kappa_{is}$ is the number of links of node $i$ to nodes in community $s$ and $k_{i}$ is the degree of the node $i$. 
And then, based on the $z$-score we identify the hubs and non-hub nodes. If $z \ge 2.5$ the nodes are classified as hubs while non-hubs are identified with $z<2.5$. The hubs and non-hubs are further classified based on the participation coefficient $P$ \cite{guimera05}. 

Non-hubs are divided into four roles :
\begin{itemize}
 \item {\it (R1) ``ultra peripheral nodes" } - Nodes with all their edges within their community ($P \le 0.05$)
 \item {\it (R2) `` peripheral nodes" } - Nodes with most of their links within their community ($0.05 < P \le 0.62$)
 \item {\it (R3) ``non-hub connector nodes" } - Nodes with many links to other community ($ 0.62 < P \le 0.80$)
  \item {\it (R4) ``non-hub kinless nodes" } - Nodes with links homogeneously distributed among all communities ($P > 0.80$)
\end{itemize}

Hubs are divided into three roles :
\begin{itemize}
 \item {\it (R5) ``provincial hubs" } - Hub nodes with majority of edges within their community ($P \le 0.30$)
 \item {\it (R6) `` connector hubs" } - Hub nodes with many connections to most of other communities ($0.30 < P \le 0.75$)
 \item {\it (R7) ``kinless hubs" } - Hub nodes with links homogeneously distributed among all communities ($ P>0.75$)
\end{itemize}

{\bf We apply the role-classification approach to the communities detected from Rosval's algorithm}. For each player in the BPNs of different teams we calculate the within-community degree $z_{i}$ and $P_{i}$. Our analysis show that there are no players who fall in the category of `R4' and `R7'.  We observe that for the Australian BPN {\it KD Walters} is the only connector hub (R6). Similarly {\it SR Tendulkar} is the only connector hub in the BPN of India. However for England we observe six  connector hubs - {\it MC Cowdrey}, {\it DI Gower}, {\it L Hutton}, {\it TG Evans}, {\it DCS Compton} and {\it WR Hammond}.  It is interesting to note that legendary Australian batsman {\it Sir DG Bradman} is not a connector hub. We observe that players falling under the category of R6 have long career and batted in various positions - batting position $3$ to $7$, allowing them to form partnership with not only the top order players but also middle order and lower order batsmen. On the other hand players like Australia's {\it Sir DG Bradman} \& {\it SR Waugh} \& {\it RT Ponting}, England's {\it IT Botham}, India's {\it SM Gavaskar}, Pakistan's {\it Javed Miandad}, West Indies'  {\it IVA Richards} batted at fixed batting positions most of the time and  belong to the R2 category. All the R2 players formed partnership mostly with players within their community. We observe that specialist openers and lower order batsmen come under R1 category, since these players always bat with fixed players. Examples include South Africa's {\it GC Smith} (an opener), {\it AA Donald} (a lower order batsman). Interestingly players like {\it JH Kallis} (South Africa), {\it VVS Laxman} (India), {\it SC Ganguly} (India), {\it BC Lara} (West Indies), {\it S Chanderpaul} (West Indies),  act as provincial hubs (R5). Incidentally R5 players are middle order batsmen who partner mostly with other contemporary middle order players, while occasionally forming partnership with lower order batsmen. Applying role-classification approach on communities detected from  Newman's algorithm we obtain a similar result - {\it MC Cowdrey}, {\it DI Gower}, {\it WR Hammond} and   {\it SR Tendulkar}  belong to `R6' category. Interestingly {\it Sir DG Bradman}, {\it SR Waugh}, {\it IT Botham}, {\it SM Gavaskar}, {\it Javed Miandad}, {\it IVA Richards} also occupy the `R2' category. Thus the roles played by batsmen in BPN provides us with an idea about the batting position of the players. Next, we extend our analysis to quantify the performance and importance of a player in batting line up.

\subsection*{Performance Index}
\begin{figure*}
\begin{center}
\includegraphics[width=10.5cm]{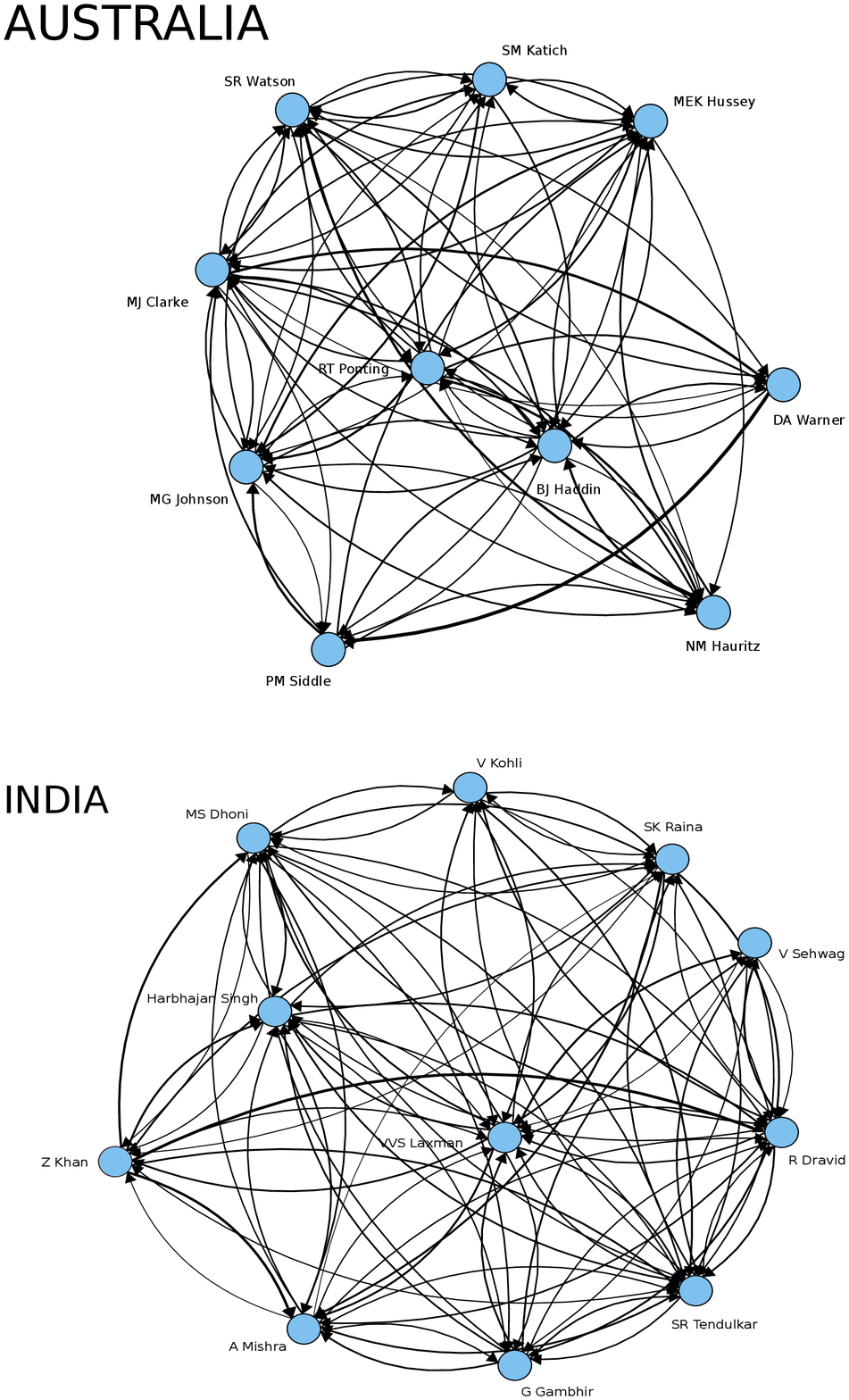}
\caption{ \label{fig:directed}(Color online) Subgraph of weighted and directed networks of batting partnership for Australia and India for Test matches played between $2009$ and $2012$. The weight of each link is equal to the ratio of runs scored by a batsman to the total number of runs scored in a batting partnership. }
\end{center}
\end{figure*}

Between $2009$ and $2012$ cricinfo made available the individual contribution in a batting partnership in Test cricket. We generate weighted and directed networks of batting partnership for all teams, where the weight of a link is equal to the fraction of runs scored by a batsman to the total runs scored in a partnership with another batsman. Thus if two batsmen $A$ and $B$ score $n$ runs between them where the individual contributions are $n_A$ and $n_B$, then a directed link of weight $\frac{n_{A}}{n}$ from $B$ to $A$. In Figure~\ref{fig:directed} we show an example of weighted and directed batting partnership network for two teams - Australia and India. The batting partnership networks are generated for all the Test matches played by the teams during $2009-2012$. We quantify the batting performance of individual players within a team by  analyzing the centrality scores - in-strength, PageRank score, betweenness centrality and closeness centrality.

For the weighted network the in-strength $s_{i}^{in}$ is defined as 
\begin{equation}
s_{i}^{in} = \displaystyle\sum_{j \ne i} \omega_{ji}
\end{equation}
where $\omega_{ji}$ is given by the weight of the directed link.

We quantify the importance or `popularity' of a player with the use of a complex network approach and evaluating the PageRank score. Mathematically, the process is described by the system of coupled equations
\begin{equation}
    p_i =  \left(1-q\right) \sum_j \, p_j \, \frac{{\omega}_{ij}}{s_j^{\textrm{out}}}
+ \frac{q}{N} + \frac{1-q}{N} \sum_j \, \delta \left(s_j^{\textrm{out}}\right) \;\; ,
\label{eq:pg}
\end{equation}
where ${\omega}_{ij}$ is the weight of a link and $s_{j}^{out}$ = $\Sigma_{i} {\omega}_{ij}$ is the out-strength of a link. $p_i$ is the PageRank score assigned to team $i$ and represents the fraction of the overall ``influence'' sitting in the steady state of the diffusion process on vertex $i$ (\cite{radicchi11}).  $q \in \left[0,1\right]$ is a control parameter that  awards a `free' popularity to each player and $N$ is the total number of players in the network. 
The term $ \left(1-q\right) \, \sum_j \, p_j \, \frac{{\omega}_{ij}}{s_j^{\textrm{out}}}$  represents the portion of the score received by node $i$ in the diffusion process obeying the hypothesis that nodes  redistribute their entire credit  to neighboring nodes. The term $\frac{q}{N}$ stands for a uniform redistribution of credit among all nodes. The term $\frac{1-q}{N} \, \sum_j \, p_j \, \delta\left(s_j^{\textrm{out}}\right)$ serves as a correction in the case of the existence nodes with null out-degree, which otherwise would behave as sinks in the diffusion process.  It is to be noted that the PageRank score of a player depends on the scores of all other players and needs to be evaluated at the same time. To implement the PageRank algorithm in the directed and weighted network, we start with a uniform probability density equal to $\frac{1}{N}$ at each node of the network. Next we iterate through  Eq.~(\ref{eq:pg}) and obtain a steady-state set of PageRank scores for each node of the network. Finally, the values of the PageRank score are sorted to determine the rank of each player. According to tradition, we use a uniform value of $q=0.15$. This choice of $q$ ensures a higher value of PageRank scores \cite{radicchi11}.

Another performance index is betweenness centrality, which measures the extent to which a node lies on a path to other nodes. In cricketing terms, betweenness centrality measures how the run scoring by  a player during a batting partnership depends on another player. Batsmen with high betweenness centrality are crucial for the team for scoring runs without losing his wicket. These batsmen are important because their dismissal has a huge impact on the structure of the network.  So a single player with a high betweenness centrality is also a weakness, since the entire team is vulnerable to the loss of his wicket. In an ideal case, every team would seek a combination of players where betweenness scores are uniformly distributed among players. Similarly the opponent team would seek to remove the player with higher betweenness centrality. 
Closeness centrality measures how easy it is to reach a given node in the network \cite{wasserman,peay}. In cricketing terms, it measures how well connected a player is in the team. Batsmen with high closeness allow the option for changing the batting order depending on the nature of the pitch or match situation. 

\begin{figure*}
\begin{center}
\includegraphics[width=14.5cm]{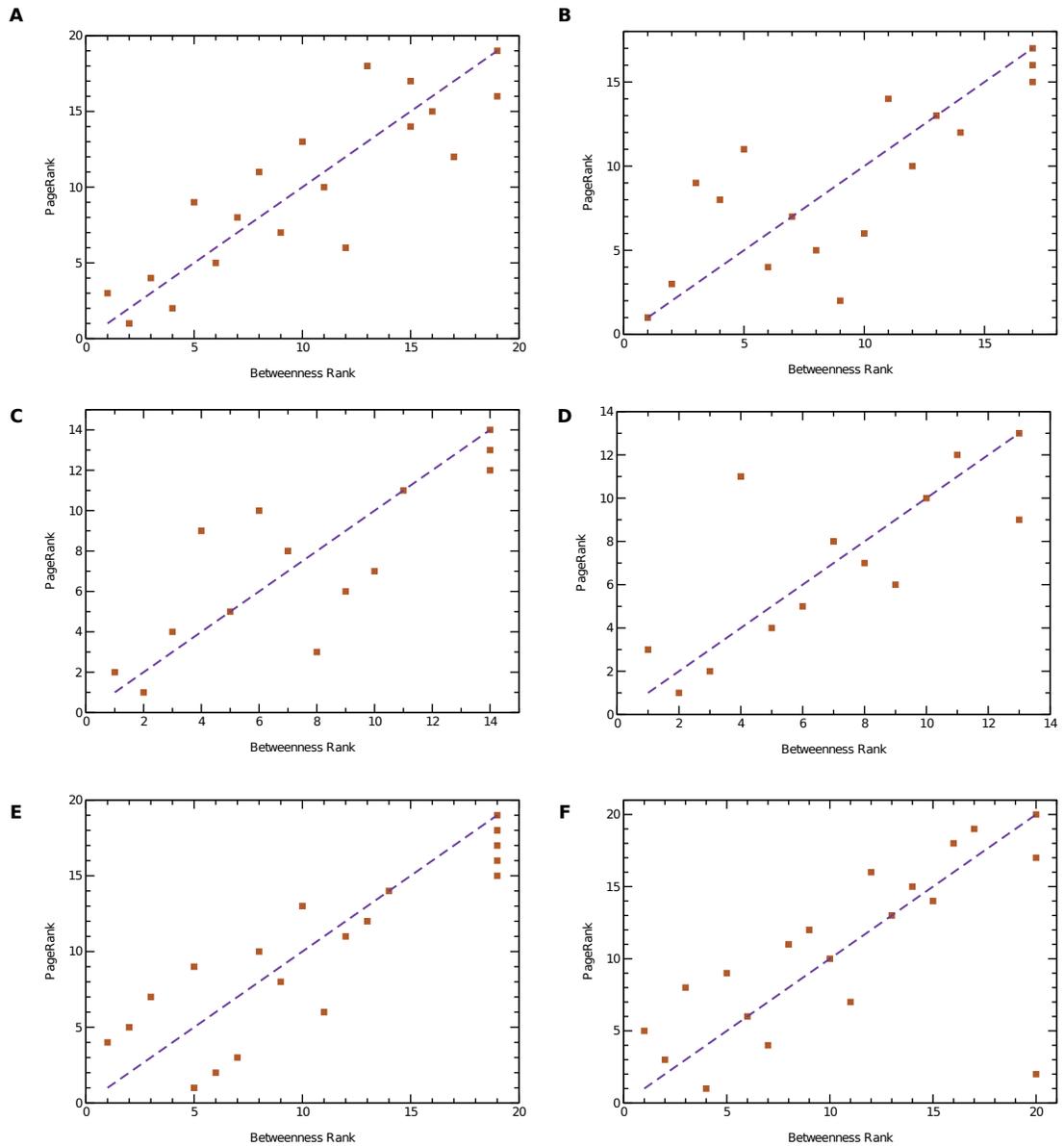}
\caption{ \label{fig:corr}(Color online) Spearman correlation coefficient ($\rho$) between the ranking based on PageRank and the one based on the betweenness centrality for (A) Australia $\rho=0.867$ (B) England $\rho=0.771$(C) South Africa $\rho=0.801$ (D) India $\rho=0.775$ (E) Pakistan $\rho=0.867$ and (F) West Indies $\rho=0.676$. }
\end{center}
\end{figure*}

 In  Table~\ref{tableS3} we compare the performance of players for different teams. According to page rank, in-strength and closeness measures {\it MJ Clarke} is the most successful batsman for Australia. This is also justified by the fact that between $2009$ and $2012$ he was the most prolific scorer for his team scoring ten centuries including a triple-hundred. Interestingly {\it MG Johnson}, who is not a specialist batsman is the most central player for Australia during the same period. During this time, batting mostly at batting position $8$, {\it MG Johnson} scored five half centuries and a century against strong opposition like South Africa. Similar arguments hold for South Africa's {\it DW Steyn} who is most central player for his team even though he bats mostly at lower position. On the other hand for the subcontinent teams the most central players are the specialist batsmen. For example {\it VVS Laxman} (India) is the most central player during $2009-2012$. During this period {\it VVS Laxman} scored nineteen half-centuries and four centuries batting mostly at fifth or sixth position along with the lower order batsmen. Thus we observe that subcontinent teams are more dependent on specialist batsmen, while for teams like Australia and South Africa, even the lower order batsmen contribute to the team's score.  In Figure~\ref{fig:corr} we show the correlation among the ranking schemes based on betweenness centrality and PageRank for different teams.
 
 Our analysis is also applicable to select the Man of the Match (MOM) after a match (or Man of the Series after a tournament). In order to validate our point, we choose the ICC World Cup Final played between India and Sri Lanka in April 2011.  We observe that Sri Lanka's {\it DPMD Jayawardene} and India's {\it G Gambhir} and {\it MS Dhoni} are the top three performers of the match based on all the centrality scores (See Table~\ref{tableS4}). All these three players were top contenders for the MOM award - {\it DPMD Jayawardene} was the highest scorer of the match, {\it G Gambhir} the highest scorer for Indian Innings and {\it MS Dhoni} won the match for India. Thus we see that tools of SNA is able to capture the consensus opinion of cricket experts. Although judged by cricket experts, {\it MS Dhoni} was named the Man of the Match, according to the centrality measures {\it DPMD Jayawardene} was the most deserving candidate.

\section*{Conclusion}
To summarize, we investigated the structural properties of batsmen partnership network (BPN) in the history of Test cricket ($1877-2012$). Our study reveals that SNA is able to examine individual level network properties among players in cricket. We observe that the networks of batsmen partnership display small-world properties and are weakly disassortative in nature. 
Similar small world phenomena is observed for bowlers as well (See Appendix).
The batting partnership networks not only provides a visual summary of proceedings of matches for various teams, they are also used to analyze the importance or popularity of a player in the team. We identify the pattern of play for various teams and potential weakness in batting line-up. Identifying the `central' player in a batting line up is always crucial for the home team as well as the opponent team. We observe that subcontinent teams depend more on the top batsmen, while other teams involve in team-game. In our analysis an interesting fact stands out - players with high batting average are not necessarily the one with highest betweenness centrality or degree. The existence of communities in the BPNs call for the definition of the role of each player. The players are classified into different roles based on the pattern of intra-community and inter-community connections. We observe that {\it Sir DG Bradman}, considered the greatest cricketer till date is not a ``connector hub". We believe this phenomena is not restricted to Cricket alone. It is important to study the the role played by other sport greats like {\it Babe Ruth} in baseball or {\it Pele} in soccer. Another possibility is to study the so called ``Shane Battier effect", where athletes makes the team better by their presence in the team. There are some additional features which could be applied in our analysis. The networks in our study are static and we assumed all the batsmen are equally athletic in the field. One could add an ``athletic index" as an attribute to each batsman. Also adding the fielders as additional nodes in the networks could provide us with a true picture of the difficulty faced by a batsman while scoring.

{\bf In real life many networks display community structure : subsets of nodes having dense node-node connections, but between which few links exist. Identifying community structure in real world networks have could help us to understand and exploit these networks more effectively. Presence of community structure could very well be due to the social groupings. We identify communities in the BPN of different countries by applying two community detection algorithm - Newman's modularity optimization algorithm and Rosval's random walk algorithm. One fact which stands out in both these algorithms is that contemporary batsmen who partner frequently with each other belong to the same community. This helps us in understanding which players are more comfortable in forming batting partnership with each other, which in turn would produce a balanced team. This can be extended well beyond Cricket in particular and sports in general. For example it was shown for United States college football that it incorporates a known community structure \cite{girvan02}. The college teams were  divided into conferences containing around $8$-$12$ teams each. It was observed that Newman and Girvan's algorithm identifies the conference structure with a high degree of success \cite{girvan02}. Even in soccer and basketball one could represent the team dynamics as a directed and weighted network, with the nodes representing players and links representing ball movements. The network structure for various teams reveal different strategies adopted by different teams. Identifying communities in such networks would reveal the various roles played by the players. Further one can also compare teams based on relative player involvement. One way to evaluate this is by quantifying the frequency of triangles \cite{milo} within the BPN. Such an approach has been adopted in identifying relative involvement of players in Basketball \cite{waters}. This can also be extended to study player involvement in Soccer as well. For example, from a fan's perspective one can study the connectedness of S\'{o}crates, Zico, Falc\~{a}o and \'{E}der in the $1982$ Brazilian soccer team. }

Our analysis thus provides a platform to quantify the over-all performance of a batsman or bowler in a tournament. For example, scoring runs against quality opposition deserves more credit than scoring runs or taking wickets against mediocre teams. To date the ranking of players has been done on the basis of batting or bowling average. Revised ranking schemes based on Gini coefficient have also been studied\cite{Vani2010}. However such ranking schemes do not take into account the quality of bowling attack. A network based approach could  address the issue of relative performance of one player against other. Potentially our study leaves a wide range of open questions which could stimulate further research in other team sports as well.

\section*{Acknowledgements}

The author thanks the cricinfo website for public availability of data. We also gratefully acknowledge helpful discussions with R. Mukogo and R. K. Pan as well. The author also thanks the two unknown referees for their comments which helped in improving the manuscript. Financial support for this research was provided by the Kellogg Graduate School of Management, Northwestern University.

\begin{figure*}
\begin{center}
\includegraphics[width=16cm]{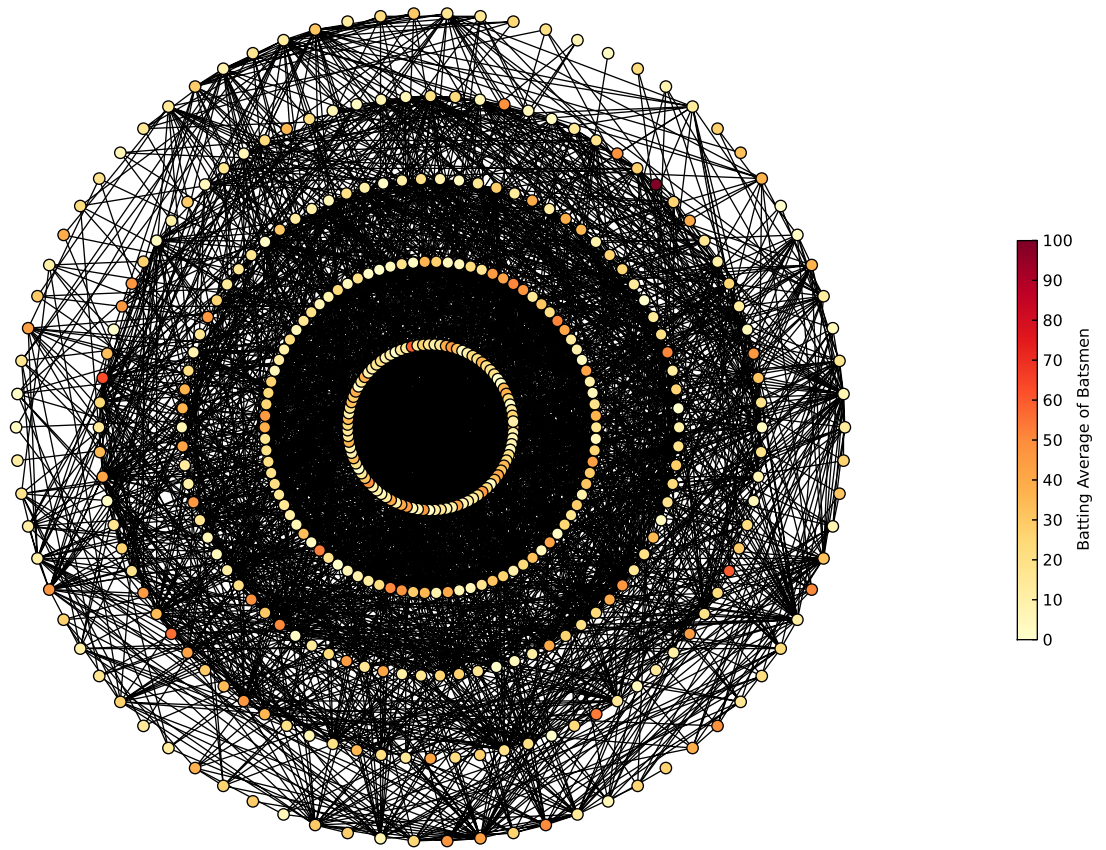}
\caption{ \label{fig:comm}(Color online) Community structure of batsmen partnership network for Australia in Test cricket ($1877-2012$). Each shell represents a community and each node indicates a player. The color of each node is proportional to the batting average of the player. Its not difficult to realize where {\it Sir Don Bradman} with a batting average of $99.94$ is located in the network. }
\end{center}
\end{figure*}

\clearpage

\begin{table}[ht!]
\centering
\caption{{ Top five batsmen ranked according to the betweenness centrality for different countries in Test cricket ($1877-2012$). }  We compare the ranks with degree and batting average of the batsman.}
\begin{tabular}{ccccc}

\hline
\multirow{2}{*}{}
 {Country} & {Betweenness} & {Degree} & {Batting Average}& {Player}\\ 
\hline 

\multirow{5}{*}{{Australia}}	
	& \multirow{6}{*}{}	
	0.414 & 35 & 48.41 & RN Harvey \\
& 0.329 & 37 & 46.81 & RB Simpson \\
& 0.310 & 35 & 99.94 & DG Bradman \\
& 0.184 & 42 & 22.65 & WAS Oldfield \\
& 0.172 & 38 & 31.83 & GM Wood \\
\hline 

\multirow{5}{*}{{England}}	
	& \multirow{6}{*}{}	
	0.236 & 15 & 8.00 & A Ward \\
& 0.145 & 51 & 32.75 & APE Knott \\
& 0.094 & 61 & 58.45 & WR Hammond \\
& 0.093 &  52 & 22.53 & JE Emburey \\
& 0.092 &  60 & 42.58 & GA Gooch \\
\hline 

\multirow{5}{*}{{South Africa}}	
	& \multirow{6}{*}{}	
	0.294 & 21 & 5.00 & JP Duminy \\
& 0.249 &  51 &40.77 & HW Taylor \\
& 0.187 & 38 & 48.88 & B Mitchell \\
& 0.119 & 25 & 43.66& EAB Rowan \\
& 0.099 & 41 & 29.78 & AW Nourse \\
\hline 

\multirow{5}{*}{{India}} 
	& \multirow{6}{*}{}
	0.306 & 33  & 32.56 & P Roy \\
& 0.304 & 48  & 31.05 & N Kapil Dev \\
& 0.179 & 64  & 55.44 & SR Tendulkar \\
& 0.122 & 51  & 51.12 & SM Gavaskar \\
& 0.101 & 52  & 17.77 & A Kumble \\

\hline 

\multirow{5}{*}{{Pakistan}}	
	& \multirow{6}{*}{}	
	0.162  & 44   & 52.57 & Javed Miandad\\
 & 0.159  & 44  & 37.69 &  Imran Khan\\
 & 0.149  & 61  & 49.60 &  Inzamam ul Haq\\
 & 0.122  & 31  & 39.17 &  Mushtaq Mohammad\\
 & 0.119  & 52   & 43.69 & Saleem Malik\\
\hline 

\multirow{5}{*}{{West Indies}}	
	& \multirow{6}{*}{}	
	0.243  & 43   & 15.50 & GA Headley\\
 & 0.187  & 37   & 6.00 & LR Gibbs\\
 & 0.180  & 30   & 40.00 & CL Walcott\\
 & 0.139  & 78  &50.44 &  S Chanderpaul\\
 & 0.138  & 38   & 42.29 & DL Haynes\\

\hline

\end{tabular}
\label{tableS1}
\end{table}

\begin{table}[ht!]
\centering
\caption{{Community structure of batsmen-partnership network for different countries in Test cricket ($1877-2012$) as done by Newman's modularity maximization algorithm. }  We provide the size of each community, the mean batting average in each community, $95\%$ CIs for each community and three prominent players in every community. $Q$ is the modularity of community structure. For each community we report the standard $z$-scores, with value of $z$ greater than $2$ reported in bold.}
\begin{tabular}{ccccc}

\hline
\multirow{2}{*}{}
 {Country} & {Size} & {Mean} & {95$\%$ CIs}& {Prominent players}\\ 
\hline 

\multirow{3}{*}{{Australia}}	
	& \multirow{6}{*}{}	
	89 & 22.84 & [  20.04, 25.68  ] & DG Bradman, SJ McCabe, WH Ponsford \\
 & 85 & 24.27 & [  21.04, 27.51  ] &  KR Miller, IR Redpath, RR Lindwall\\
Q=0.69 & 	79 & 22.61 & [  19.64, 25.68  ] & C Bannerman, FA Iredale, VT Trumper\\
 $z$-score = {\bf 103} & 	83 & 16.09 & [  13.78, 18.48  ] & SR Waugh, RT Ponting, AC Gilchrist\\
 & 	77 & 23.58 & [  19.78, 27.63  ] & GS Chappell, AR Border, DC Boon\\
\hline 

\multirow{3}{*}{{England}}	
	& \multirow{6}{*}{}	
	154 & 18.66 & [  16.73, 20.67  ] & WG Grace, W Barnes, G Ulyett\\
& 129 & 21.96 & [  19.37, 24.65  ] & GB Legge, CAG Russell, EH Hendren\\
 Q=0.69 & 127 & 21.26 & [  18.87, 23.81  ] & JC Laker, JDB Robertson, PBH May\\
$z$-score={\bf 113} & 121 & 19.02 & [  16.78, 21.39 ] & N Hussain, MA Atherton, GA Hick\\
& 102 & 22.34 & [  20.08, 24.67  ] & JH Edrich, IT Botham, JM Brearley\\
\hline 

\multirow{3}{*}{{South Africa}}	
	& \multirow{6}{*}{}	
	61 & 20.91 & [  17.45, 24.64  ] & TL Goddard, EJ Barlow, KC Bland\\
& 37 & 19.63 & [  14.9 , 24.86 ] & GM Fullerton, HF Wade, AD Nourse\\
Q=0.67& 61 & 15.02 & [  12.35, 17.77  ] & GA Faulkner, HW Taylor, JW Zulch\\
$z$-score = {\bf 63.69}& 44 & 7.57 & [  5.66, 9.587  ] & JH Sinclair, GK Glover, WH Milton \\
& 26 & 18.73 & [  13.51, 24.24  ] & RH Catterall, JAJ Christy, JFW Nicholson\\
& 73 & 22.74 & [  19.58, 25.98  ] & WJ Cronje, JH Kallis, SM Pollock\\

\hline 

\multirow{3}{*}{{India}} 
	& \multirow{6}{*}{}
	59 & 18.10 & [  14.84, 21.62  ] &  VS Hazare, HM Mankad, P Roy\\
	& 32 & 19.49 & [  14.84, 24.56  ] & L Amarnath, S Mushtaq Ali,  CK Nayudu\\
Q = 0.58	& 53 & 20.25 & [  17.14, 23.46  ] & SM Gavaskar, GR Viswanath, M Amarnath\\
$z$-score = {\bf 67} 	& 37 & 21.51 & [  17.05, 26.21  ] & M Azharuddin, N Kapil Dev, DB Vengsarkar\\
	& 76 & 19.95 & [  16.88, 23.07  ] & VVS Laxman, SR Tendulkar, MS Dhoni\\
\hline 

\multirow{1}{*}{{Pakistan}}	
	& \multirow{6}{*}{}	
	50 & 18.40 & [  14.88, 22.12  ] & Wasim Akram, Inzamam ul Haq, Yousuf Youhana\\
Q=0.53& 42 & 23.45 & [  19.22, 27.74  ] & Shoaib Malik, Younis Khan, Salman Butt\\
$z$-score={\bf 47.86}& 51 & 19.96 & [  16.68, 23.38  ] & Hanif Mohammad, Fazal Mahmood, Aftab Baloch \\
& 53 & 23.79 & [  19.79, 27.87  ] & Javed Miandad, Imran Khan, Zaheer Abbas \\
\hline 

\multirow{3}{*}{{West Indies}}	
	& \multirow{6}{*}{}	
	45 & 21.16 & [  16.38, 26.66  ] & GA Headley, JB Stollmeyer, LN Constantine\\
& 77 & 23.06 & [  18.99, 27.56  ] & FMM Worrell, ED Weekes, GS Sobers\\
Q=0.62& 57 & 24.20 & [  20.80, 27.68  ] & IVA Richards, CH Llyod, DH Haynes \\
$z$-score = {\bf 70.67}& 58 & 19.78 & [  16.85, 22.94  ] & S Chanderpaul, R Sarwan, CH Gayle\\
& 48 & 17.62 & [  14.17, 21.11  ] & BC Lara, CL Hooper, RB Richardson \\

\hline

\end{tabular}
\label{tableS2}
\end{table}

\begin{table}[ht!]
\centering
\caption{{Community structure of batsmen-partnership network for different countries in Test cricket ($1877-2012$) as done by Rosval's  algorithm. }  We provide the size of each community, the mean batting average in each community, $95\%$ CIs for each community and three prominent players in every community.}
\begin{tabular}{ccccc}

\hline
\multirow{2}{*}{}
 {Country} & {Size} & {Mean} & {95$\%$ CIs}& {Prominent players}\\ 
\hline 

\multirow{6}{*}{{Australia}}	
	& \multirow{6}{*}{}	
 108 & 19.62 &  [  16.96,  22.63  ]  & SR Waugh, RT Ponting, AC Gilchrist\\
& 74 & 22.66  &  [  19.90,  25.79  ]  & RN Harvey, KR Miller, RR Lindwall, \\
& 100 & 22.78  &  [  19.53,  26.23  ]  & DG Bradman, SJ McCabe, WH Ponsford \\
& 42 & 19.80 &  [  16.30,  23.44  ]  & KJ Hughes, GM Wood, RB Simpson \\
& 55 & 24.12  &  [  20.36,  28.13  ]  & AR Border, DC Boon, CJ McDermott \\
& 34 & 23.60 &  [  19.19,  27.95  ]  & MEK Hussey , MJ Clarke, MG Johnson\\
\hline 

\multirow{8}{*}{{England}}	
	& \multirow{6}{*}{}	
	
 5  & 17.37  &  [   4.12,  33.25  ]  & BAF Grieve, Hon. CJ Coventry, AJ Fothergill\\
 & 5 &  22.75  &  [  14.65,   32.25  ]  & W Chatterton, WL Murdoch, A Hearne \\
 & 165 &  23.50 &  [  21.34,   25.76  ]  & GB Legge, EH Hendren, G Boycott\\
& 132  & 19.30 &  [  17.48,   21.43 ]  & GA Gooch, IT Botham, JM Brearley \\
 & 131  & 18.41 &  [  16.35,   20.54  ]  & MA Atherton, N Hussain, GA Hick \\
 & 61  & 21.19  &  [  17.77,   25.06  ]  & A Flintoff, MP Vaughan, AF Giles \\
 & 115  & 19.65 &  [  17.27,   22.24  ]  & CAG Russell, PBH May, JC Laker  \\
 & 6  & 17.98 &  [  11.71,   24.24  ]  & CP Buckenham, HDG Leveson Gower, NC Tufnell \\
 & 8  & 19.36 &  [   8.32,   33.12  ]  & Lord Hawke, AJL Hill, CW Wright \\
 & 5  & 22.262  &  [  8.06,   36.32  ]  & FW Milligan, AE Trott, AG Archer \\

\hline 

\multirow{8}{*}{{South Africa}}	
	& \multirow{6}{*}{}	
	 117  &  21.98 &  [   19.26,   24.84  ]  &  WJ Cronje, JH Kallis, SM Pollock\\
&  57 &  20.96 &  [  16.82,   25.58  ]  & KC Bland, EJ  Barlow, HW Taylor\\
&  66  & 16.67  &  [  14.44,  19.33  ]  & JH Sinclair, JW Zulch, GA Faulkner\\
&  19  & 6.43 &  [  3.82,   9.80  ]  & CH Mills, CH Vintcent, G Cripps \\
&  27  & 9.67 &  [   7.40,   12.22  ]  & J Middleton, GA Rowe, GM Fullerton\\
&  3  & 8.72  &  [  3.5,   19.0  ]  & GF Bissett, AL Ochse, HLE Promnitz\\
&  11  & 9.09  &  [  4.60,   14.07  ]  & C Newberry ,  HW Chapman , JL Cox\\
&  2  & 10.87 &  [   0.5,   21.25  ]  & D Taylor, FL le Roux\\

\hline 

\multirow{5}{*}{{India}} 
	& \multirow{6}{*}{}
 78  & 19.94 &  [  16.94,   23.03  ]  & SR Tendulkar, R Dravid, VVS Laxman\\
&  59 &  18.10  &  [  14.69,   21.61  ]  & VS Hazare, HM Mankad, P Roy \\
&  56  & 22.46 &  [  18.87,   26.19  ]  & SM Gavaskar, M Azharuddin, M Amarnath \\
&  32  & 19.49 &  [  15.05,   24.42  ]  & L Amarnath, S Mushtaq Ali,  CK Nayudu\\
&  32  & 17.89 &  [  13.98,   21.72  ]  & GR Viswanath, S Venkataraghavan, Nawab of Pataudi jnr\\
\hline 

\multirow{3}{*}{{Pakistan}}	
	& \multirow{6}{*}{}	
	51  & 23.00 &  [  19.05,   27.09  ]  & Younis Khan, Yousuf Youhana, Shoaib Malik \\
& 48  & 18.47 &  [  14.92,   22.33  ]  & Inzamam ul Haq, Saleem Malik, Wasim Akram \\
&  52  & 24.86  &  [  20.90,   28.93  ]  & Javed Miandad, Imran Khan,  Zaheer Abbas \\
&  45  & 18.47 &  [  15.50,   21.88  ]  & Hanif Mohammad, Intikhab Alam, Fazal Mahmood \\
\hline 

\multirow{5}{*}{{West Indies}}	
	& \multirow{6}{*}{}	
	 104 & 18.64  &  [  16.37,   20.86  ]  & S Chanderpaul, BC Lara, CL Hooper\\
&  40 &  23.42  &  [  18.44,   29.36  ]  & GS Sobers, RB Kanhai, LR Gibbs\\
&  52  & 24.25 &  [   20.71,   28.05 ]  & CH Lloyd, DL Haynes,  IVA Richards\\
&  45  & 21.16 &  [  16.50,   26.64  ]  & GA Headley, LN Constantine, GC Grant \\
&  44  &  22.98 &  [  17.47,   29.46  ]  & FMM Worrell, ED Weekes, JB Stollmeyer\\

\hline

\end{tabular}
\label{tableS6}
\end{table}

\begin{table}[ht!]
\centering
\caption{{Performance of batsmen for different countries in Test cricket between $2009$ and $2012$. } The top five performers are ranked according to their PageRank score and their performance is compared with In-strength, betweenness centrality and closeness centrality.}
\begin{tabular}{cccccc}

\hline
\multirow{2}{*}{}
 {Country} & {Players} & {PageRank} & {In-strength}& {Betweenness} & {Closeness}\\ 
\hline 

\multirow{6}{*}{{Australia}}	
	& \multirow{6}{*}{}	
	MJ Clarke & 0.0693 & 12.3688 & 0.1372  & 0.7307\\
 & BJ Haddin & 0.0664  & 12.1215 &  0.1187 & 0.7037\\
& MG Johnson & 0.0619 & 12.3393  & 0.1863 & 0.7169\\
 & MEK Hussey & 0.0593 & 11.8637  & 0.1209 & 0.7037 \\
 & PM Siddle & 0.0460 & 8.4558  & 0.0711  & 0.6551 \\
\hline 

\multirow{6}{*}{{England}}	
	& \multirow{6}{*}{}	
	MJ Prior & 0.0870 & 12.0851 & 0. 1894 & 0.8181\\
 & GP Swann & 0.0712  & 11.0006 &  0.0512 & 0.7105\\
& KP Pietersen & 0.0653 & 9.1076  & 0.1054 & 0.6923\\
& IR Bell & 0.0619 & 8.5239  & 0.0683 & 0.7105 \\
 & IJL Trott & 0.0593 & 7.4103  & 0.0569  & 0.7297 \\
\hline 

\multirow{6}{*}{{South Africa}}	
	& \multirow{6}{*}{}	
	AB de Villiers & 0.0914 & 8.9988 & 0.1205  & 0.7666\\
 & DW Steyn & 0.0808  & 7.7999 &  0.2154 & 0.7666\\
& JH Kallis & 0.0769 & 7.9980  & 0.0573 & 0.7187\\
 & MV Boucher & 0.0768 & 7.7552  & 0.1146 & 0.7419 \\
 & M Morkel & 0.0732 & 7.5571  & 0.1047  & 0.7187 \\

\hline 

\multirow{6}{*}{{India}} 
	& \multirow{6}{*}{}
	MS Dhoni & 0.0847 & 13.8579 & 0.2075  & 0.7948\\
 & SR Tendulkar & 0.0825  & 11.4711 &  0.1204 & 0.7045\\
& VVS Laxman & 0.0800 & 10.9430  & 0.2139 & 0.7750\\
 & R Dravid & 0.0722 & 8.9096  & 0.0784 & 0.6595 \\
 & Harbhajan Singh & 0.0573 & 9.2165  & 0.0365  & 0.6326 \\
\hline 

\multirow{6}{*}{{Pakistan}}	
	& \multirow{6}{*}{}	
	Umar Akmal & 0.0656 & 12.8386 & 0.0754  & 0.6667\\
 & Younis Khan & 0.0575  & 9.2505 &  0.1460 & 0.6315\\
& Misbah ul Haq & 0.0543 & 10.1389  & 0.0674 & 0.6545\\
 & Kamran Akmal & 0.0540 & 9.8195  & 0.0634 & 0.6101 \\
 & Umar Gul & 0.0536 & 11.2592  & 0.1873  & 0.72 \\
\hline 

\multirow{6}{*}{{West Indies}}	
	& \multirow{6}{*}{}	
	SJ Benn & 0.0693 & 15.5136 & 0.0940  & 0.5833\\
 & JE Taylor & 0.0664  & 16.1434 &  0.0 & 0.4516\\
& S Chanderpaul & 0.0619 & 10.9692  & 0.1202 & 0.6774\\
 & DJG Sammy & 0.0593 & 11.0091  & 0.0784 & 0.6363 \\
 & TW Dowlin & 0.0460 & 9.6061  & 0.1945  & 0.6176 \\

\hline

\end{tabular}
\label{tableS3}
\end{table}

\begin{table}[ht!]
\centering
\caption{{Performance of batsmen during the ICC 2011 World Cup final played between Sri Lanka and India. }  The players are ranked according to their PageRank score and the performance is compared with In-strength, betweenness centrality and closeness centrality.}
\begin{tabular}{cccccc}

\hline
\multirow{2}{*}{}
 {Country} & {Players} & {PageRank} & {In-strength}& {Betweenness} & {Closeness}\\ 
\hline 

\multirow{6}{*}{{Sri Lanka}}	
	& \multirow{6}{*}{}	
	DPMD Jayawardene & 0.3651 & 2.3751 & 0.8571  & 0.7\\
 & KC Sangakkara & 0.1537  & 0.9096 &  0.4761 & 0.5833\\
& NLTC Perera & 0.1237 & 0.8461  & 0.0 & 0.4375\\
 & TM Dilshan & 0.1146 & 1.1942  & 0.2857 & 0.4375 \\
 & KMDN Kulasekara & 0.0789 & 0.4848  & 0.0 & 0.4375 \\
 & TT Samaraweera & 0.0644 & 0.3684 & 0.0  &0.4375\\
 & CK Kapugedera & 0.0601  & 0.3333 &  0.0 & 0.4375\\
& WU Tharanga & 0.0392 & 0.1176 & 0.3181 & 0.3181\\
 
\hline 

\multirow{4}{*}{{India}}	
	& \multirow{6}{*}{}	
	G Gambhir & 0.3549 & 1.2876 & 0.8333 & 0.8\\
 & MS Dhoni & 0.2533  & 1.1245 &  0.5 & 0.6667\\
& SR Tendulkar & 0.1428 & 0.5806  & 0.0 & 0.5\\
& Yuvraj Singh & 0.1368 & 0.3888  & 0.0 & 0.44 \\
 & V Kohli & 0.1119 & 0.4216  & 0.0  & 0.5 \\
\hline 

\end{tabular}
\label{tableS4}
\end{table}

\clearpage
\section*{Appendix}

We observed small world behavior in batting partnership network. It would be interesting to see if such small world phenomena exists among the bowlers as well. However unlike the batsmen, a bowler does not form partnership with another bowler. The game of Cricket can be represented as two-mode networks of interaction of batsmen ($B_{a}$) and bowlers ($B_{o}$). Every node in $B_{a}$ has one-to-one connection with every node in $B_{o}$. 

We define a criteria by which the bowlers are linked. 
From the two-mode network of bowlers and batsmen we construct networks composed exclusively of  bowlers. {\bf Please note that the network of bowlers is derived from the bi-partite connections of batsmen-bowlers. This is different from the batting partnership network (BPN) which is constructed between batsmen exclusively from the information of partnership among them. } 
The proceedings of a cricket match are stored in the score cards which contains the information of two competing teams, results of a match, runs scored by batsmen and wickets taken by bowlers. From the score cards available for Test cricket (from $1877$ to $2012$) we collect the information of dismissal of batsmen. A Bowler Dismissal Network (BoDN) is achieved if two bowlers dismiss the same batsman. Similar approach has been adopted in case of collaboration networks \cite{ramasco,uzzi}, actor $-$ movies \cite{amaral}, director $-$ firm \cite{davis}, scientist $-$ paper \cite{newmansp} and also soccer $-$ clubs \cite{onody04}. For Test cricket the BoDN has $3082$ nodes and $490355$ edges with a high clustering coefficient $C=0.73$, whereas its randomization yields $C=0.10$. This shows that bowlers display small-world phenomenon.

\clearpage
%

\begin{flushleft}
{\Large
\textbf{Supporting Information}
}
\\
\bf {Satyam Mukherjee}$^{1}$
\\
{1} Kellogg School of Management, Northwestern University, Evanston, IL, United States of America
\\

\end{flushleft}

\section*{Game of Cricket}
Cricket is a bat-and-ball game played between two teams of $11$ players each. The team batting first tries to score as many runs as possible, while the other team bowls and fields, trying to dismiss the batsmen. At the end of an innings, the teams switch between batting and fielding.  The International Cricket Council (ICC) is the government body which controls the cricketing events around the globe. Although ICC includes $120$ member countries, only ten countries with `Test' status - Australia, England, India, South Africa, New Zealand, West Indies, Bangladesh, Zimbabwe, Pakistan and Sri Lanka play the game extensively. There are three versions of the game - `Test', One Day International (ODI) and Twenty20 (T20) formats. Test cricket is the longest format of the game dating back to $1877$. Usually it lasts for five days involving $30-35$ hours. Shorter formats, lasting almost $8$ hours like ODI started in $1971$ and during late $2000$ ICC introduced the shortest format called T20 cricket which lasts approximately $3$ hours.  

\subsection*{Batting Partnership and Cricket Terms}
In this section we explain the concept of forming batting partnership and also the terms like "openers" and "lower-oder" batsmen. 
In cricket two batsmen always play in partnership, although only one is on strike at any time. The partnership of two batsmen comes to an end when one of them is dismissed or at the end of an innings. In cricket there are $11$ players in each team. The batting pair who start the innings is referred to as opening-pair. For example two opening batsmen $A$ and $B$ start the innings for their team. In network terminology, this can be visualized as a network with two nodes $A$ and $B$, the link representing the partnership between the two players. Now, if batsman $A$ is dismissed by a bowler, then a new batsman $C$ arrives to form a new partnership with batsman $B$. This batsman $C$ thus bats at position number three. Batsmen who bat at positions three and four are called top-order batsman. Those batsmen who bat at positions  five to seven are called middle-order batsmen. Finally the batsmen batting at position number eight to eleven are referred to as tail-enders. 
Thus a new node $C$ gets linked with node $B$. In this way one can generate an entire network of batting-partnership till the end of an innings. The innings come to an end when all $11$ players are dismissed or when the duration of play comes to an end. The score of a team is the sum of all the runs scored during a batting partnership. In our work we considered the network of batting partnership for every teams during the period $1877-$August $2012$.

\renewcommand{\thefigure}{S\arabic{figure}}

\renewcommand{\thetable}{S\arabic{table}}

\begin{table}[ht!]
\centering
\caption{{Community structure of batsmen-partnership network for different countries in Test cricket ($1877-2012$) as done by OSLOM RW algorithm. }  We provide the size of each community, the mean batting average in each community, $95\%$ CIs for each community and three prominent players in every community.}
\begin{tabular}{ccccc}

\hline
\multirow{2}{*}{}
 {Country} & {Size} & {Mean} & {95$\%$ CIs}& {Number of Singletons}\\ 
\hline 

\multirow{16}{*}{{Australia}}	
	& \multirow{6}{*}{}	
33  & 24.16  &  [  19.09,   29.53  ]  & \\
&  37 &  21.21  &  [  16.96,   25.39  ]  & \\
&  26 &  23.41 &  [  16.28,   31.51  ]  & \\
& 16  & 15.34 &  [  10.76,   20.9  ]  & \\
&  16  & 16.24  &  [  10.96,   22.53  ]  & \\
&  14  & 28.40 &  [  19.95,   37.26  ]  & \\
&  29  & 20.86  &  [   15.19,   27.55 ]  &  47\\
& 11  & 27.59  &  [  19.11,   34.30  ]  & \\
&  19  & 24.89 &  [  20.56,   29.70  ]  & \\
& 17  & 21.97  &  [  15.67,   28.36  ]  & \\
&  37  & 24.87  &  [   20.61,   29.18  ]  & \\
&  34  & 19.27  &  [  15.11,   23.38  ]  & \\
&  37 &  22.11  &  [  17.12,   26.74  ]  & \\
&  17  & 15.76  &  [  10.01,   22.83 ]  & \\
&  25  & 16.83 &  [  12.09,   21.84  ]  & \\
\hline 

\multirow{16}{*}{{England}}	
	& \multirow{6}{*}{}	
 5 &  33.77  &  [  24.36,   43.18  ]  & \\
&  4  & 18.23 &  [   3.5,   32.96  ]  & \\
&  57 &  20.78  &  [  17.56,   24.24  ]  & \\
&  45  & 23.00 &  [  19.47,   27.054  ]  & \\
& 15  & 19.27  &  [  11.51,   27.19  ]  & \\
&  47 &  21.73  &  [   17.47,   26.04  ]  & \\
&  55  & 17.61  &  [  14.48,   20.88  ]  & \\
& 30  & 21.82  &  [  17.05,   26.57  ]  & \\
&  13  & 24.12 &  [  16.78,   31.57 ]  &  87\\
&  47  & 23.94 &  [  19.65,   28.18  ]  & \\
&  32  & 18.12 &  [  13.72,   23.27  ]  & \\
& 39  & 23.40  &  [  18.90,   27.89  ]  & \\
&  27  & 20.88 &  [  15.83,   26.06  ]  & \\
&  47  & 18.30&  [  14.730,   22.03  ]  & \\
& 42  & 17.60  &  [  14.08,   21.41  ]  & \\
&  29  & 22.21  &  [  17.06,   27.88  ]  & \\

\hline 

\multirow{8}{*}{{South Africa}}	
	& \multirow{6}{*}{}	
 24  & 22.92 &  [  16.59,   29.93  ]  & \\
&  41  & 21.97  &  [  17.79,   26.63 ]  & \\
&  29  & 21.73  &  [  17.00,   26.85  ]  & \\
&  16  & 16.46  &  [  11.61   21.84  ]  & 33 \\
&  37  & 17.03  &  [  11.78,   22.44  ]  & \\
&  32  & 18.79 &  [  14.07,   23.99  ]  & \\
&  29  & 15.09  &  [  10.96,   20.28  ]  & \\
&  34  & 14.67 &  [  10.56 ,  18.67  ]  & \\

\hline

\end{tabular}
\label{tableS10}
\end{table}

\begin{table}[ht!]
\centering
\caption{{Community structure of batsmen-partnership network for different countries in Test cricket ($1877-2012$) as done by OSLOM RW algorithm Contd. }  We provide the size of each community, the mean batting average in each community, $95\%$ CIs for each community and three prominent players in every community.}
\begin{tabular}{ccccc}

\hline
\multirow{2}{*}{}
 {Country} & {Size} & {Mean} & {95$\%$ CIs}& {Number of Singletons}\\ 
\hline

\multirow{8}{*}{{India}} 
	& \multirow{6}{*}{}
12  & 14.52&  [   9.53,   20.38  ]  & \\
&  24  & 20.40  &  [   14.68,   26.17  ]  & \\
& 22  & 26.66  &  [  20.93,   32.85 ]  & \\
&  26  & 15.94 &  [  12.46,   19.39  ]  & 35\\
& 30  & 22.53  &  [  18.19,   27.44  ]  & \\
&  41  & 19.99 &  [  15.14,   24.89  ]  & \\
&  71  & 19.30  &  [  16.54,   22.34  ]  & \\
\hline 

\multirow{7}{*}{{Pakistan}}	
	& \multirow{6}{*}{}	
 16  & 16.89  &  [  12.70,   21.12  ]  & \\
& 19 &  17.19  &  [  11.45,   23.03 ]  & \\
&  24 &  27.19  &  [  21.91,   32.20  ]  & 25\\
& 38  & 21.01  &  [  16.88,   25.85  ]  & \\
&  45  & 20.95 &  [  17.00,   25.02  ]  & \\
& 31  & 23.16 &  [   18.51,   27.90  ]  & \\
\hline 

\multirow{8}{*}{{West Indies}}	
	& \multirow{6}{*}{}	
 47  & 23.14 &  [  17.90,   28.82 ]  & \\
&  34 &  21.43  &  [  16.58,   26.84 ]  & \\
&  16  & 13.95 &  [  9.48,   19.09  ]  & \\
&  11  & 19.57 &  [  14.27,   24.80  ]  & \\
&  49  & 26.41  &  [   21.27,   32.32  ]  & 23\\
&  48  & 18.67  &  [  15.44,   22.12  ]  & \\
&  23  & 21.48  &  [  17.09,   26.19  ]  & \\
&  36  & 19.75  &  [  15.83,   24.16  ]  & \\

\hline

\end{tabular}
\label{tableS11}
\end{table}

\clearpage

\section*{Information about players} 

\begin{itemize}

\item {\bf RN Harvey} is one of Australia's all-time favorite cricketing talents. He was a gifted left-hand batsman, brilliantly athletic fielder, and occasional off-spin bowler. He averages $48.41$ in Test cricket. 

\item {\bf RB Simpson} is a key figure in Australian cricket for more than four decades, as cricketer, captain, coach and commentator. He averages $46.81$ in Test cricket with a highest score of $311$. 

\item {\bf  DG Bradman} of Australia was the greatest Test batsman ever lived. He averages $99.94$ in Test cricket, a record which has never been eclipsed since his retirement in $1948$. 

\item {\bf WAS Oldfield} of Australia was a lower order batsman who averaged $22.65$ in Test cricket. He was also a wicketkeeper  with $78$ catches and $52$ stumpings to his credit. 

\item {\bf GM Wood} is one of Australia's finest opener averaging $31.83$ in Test cricket. His major Test highlights includes a century in the 1980 Centenary Test against England  and $111$ to set up victory in the low-scoring First Test against New Zealand in Brisbane in 1980-81.

\item {\bf} SR Waugh of Australia is the ultimate evolved cricketer. He led Australia in 15 of their world-record 16 successive Test victories. He turned Australia's form around so completely in the 1999 World Cup that they won it, and he became the first Australian to win the trophy twice. In $168$ matches he scored more that $10000$ runs at an average of $51.06$. 

\item {\bf RT Ponting} grew into Australia's most successful run-maker and only sits below Bradman in the country's overall ratings. He became the most successful captain in Test history after passing Steve Waugh's 41 wins in the 2009-10 Boxing Day Test. In the same match he overtook Shane Warne's 92 victories as the most by an individual, and he led Australia to 34 consecutive undefeated World Cup games. His batting average is $51.85$

\item {\bf MG Johnson}  is a left-handed bowler of Australia who has claimed more than $200$ wickets in $50$ Test matches he has played. Apart from this he scored a $100$ and a $96$ against strong South African team. He also score seven half-centuries.

\item {\bf A Ward} is one of the fast bolwers for England with a batting average of $8.00$, who batted as a lower order batsman. He has a high betweenness centrality since he formed batting partnership with many top order batsmen. 

\item {\bf APE Knott} played in $95$ Test matches for England with five Test hundreds to his name. He was a genuine all-rounder. In only his fourth Test, at Georgetown in $1967$, he scored $73$ not out and helped his team save the match.

\item {\bf WR Hammond} of England is regarded by all experts as one of the few batsmen par excellence. He scored $7249$ runs in $85$ matches with an average of $58.45$ and highest score of $336$ not out. Cricket experts have ranked him at par with Sir DG Bradman of Australia.

\item {\bf JE Emburey} is regarded as one of the finest spin bowlers of England, who batted in lower position. His batting average is $22.53$ and bowling average is $38.40$.

\item {\bf GA Gooch}  played in $118$ Test matches for England and scored $8900$ runs with an average of $42.58$. Overall he scored twenty centuries with the highest score of $333$. He was an enigmatic batsman and a prolific scorer in modern generation Cricket.

\item {\bf MC Cowdrey}  was the most durable English player in an era of outstanding English batsmen, with a Test career spanning more than two decades. He averaged $44.06$ scoring $22$ centuries and $7624$ runs for his team.

\item {\bf DI Gower} was a  left-handed English batsman who scored more than $8000$ runs in Test cricket with an average of $44.25$, including $18$ centuries. He was also a successful England captain.

\item {\bf LEG Ames} played for England and has been regarded as the greatest wicketkeeper-batsman the game has so far produced. His batting average is $40.56$. 

\item {\bf IT Botham} was England's greatest all-rounder. In the 102 Test matches he played, in batting he scored $5200$ runs and in bowling he took $383$ wickets. 
 
\item {\bf AJ Stewart} was opening batsman for England. In the $133$ Test matches he played, he scored $8463$ runs at an average of $39.54$.

\item {\bf JP Duminy} is one of the fast bolwers for England with a batting average of $8.00$, who batted as a lower order batsman. He has a high betweenness centrality since he formed batting partnership with many top order batsmen. 

\item {\bf HW Taylor} played in $42$ Test matches for South Africa with seven Test hundreds to his name. He was a great batsman on the matting pitches of South Africa, England and Australia. He played over a span of twenty years for his country averaging $40.77$ in  Test cricket.

\item {\bf B Mitchell} appeared in every one of South Africa's $42$ Tests from $1929$ to $1949$ and ranks among the finest batsmen produced by his country scoring eight centuries  and averaging $48.88$ in Test cricket.

\item {\bf EAB Rowan} played in $26$ Test matches for South Africa with an average of $43.66$ with the highest score of $236$. He was always regarded as a fearless player since sometimes he batted without gloves.

\item {\bf AW Nourse }  is considered a great all-rounder for South Africa - he was a dogged left-hander, seam bowler, and brilliant slip fielder, he made $15$ Test fifties but only one hundred, $111$ against Australia.

\item {\bf JH Sinclair}  did more than anyone to put South African cricket on the map. He was strong hitter of the ball and bowled fast. He scored three hundreds and claimed $63$ wickets in the $25$ Test matches he played. 

\item {\bf GC Smith} was handed the reins of captaicny at an young age of $22$ - which made him his South Africa's youngest captain. He is one of the most successful Test captains and also a good left-handed batsman, having scored more than $8000$ runs in $109$ Test matches. 

\item {\bf AA Donald} was first bowler of South Africa to claim $300$ Test wickets. He is regarded as one of the fastest bowlers in cricketing arena. 

\item {\bf JH Kallis} is one of the finest all-rounder in world cricket. Playing for South Africa he has scored more than $130000$ Test runs in $162$ matches including a hopping $44$ centuries. 

\item {\bf DW Steyn} became the fastest South African, and the 15th fastest overall, to reach 100 Test wickets in the year 2008. That same year, Steyn was named ICC Test Player of the Year after taking 86 wickets in 14 matches at an average of 18.10.

\item {\bf P Roy} played $43$ Test matches for India scoring $2442$ runs with the highest score of $173$. He was an opening batsman for India and was involved in a world-record opening stand of $413$.

\item {\bf  N Kapil Dev} is regarded as  the greatest pace bowler India has produced, and also greatest fast-bowling all-rounder. He was voted India's Cricketer of the Century during $2002$, ahead of Sunil Gavaskar and Sachin Tendulkar. His greatest feats were to lead India to the $1983$ ICC Cricket World Cup, and to take the world-record aggregate of $413$ Test wickets.

\item {\bf SR Tendulkar} has appeared in $194$ Test matches for India scoring world record of $15645$ runs which includes the  record of $51$ centuries. He has been the most complete batsman of his time, the most prolific run maker of all time, and arguably the biggest cricket icon the game has ever known.

\item {\bf SM Gavaskar} was one of the greatest opening batsmen of all time, and certainly the most successful. Playing in $125$ Tests for India he was the first to get $10000$ Test runs and $30$ centuries.

\item {\bf A Kumble }  won more Test matches for India than any other bowlers in history of Test Cricket. He played in $132$ Test matches for India as a bowler. In batting he averages $17.77$ with highest score being $110$. 

\item {\bf VVS Laxman} is best known for the majestic $281$ runs he scored against a strong Australian team in $2001$.  In $2010$, batting with a runner due to back spasms, he conjured up a magical unbeaten $73$ in a thrilling run-chase against Australia.

\item {\bf SC Ganguly} of India is placed sixth by Wisden in their greatest ODI batsmen of all time. He proved to be a tough, intuitive and uncompromising leader and galvanized his team to win abroad in Test matches and also led to the World Cup final in $2003$. He is one of most successful captains of India.

\item {\bf MS Dhoni} is one of the most successful captains of not only India but also the world. He is a great ODI batsman and averages $51.85$.

\item {\bf G Gambhir} is an opening batsman for India. In his $147$ ODIs , he scored $11$ centuries. He also scored $9$ centuries - centuries to set up wins, centuries to bat opposition out, and centuries to hold on for draws, including the near 11-hour marathon in Napier. He scored a match-winning 97 in the 2011 World Cup final. 

\item {\bf DPMD Jayawardene} is a  prolific, elegant and utterly classy batsman with a huge appetite for runs, and a calm yet authoritative captain of Sri Lankan team.  With over 10,000 runs in both Tests and ODIs - and a captaincy stint that included a World Cup final appearance - it can safely be said that he has met that challenge more than adequately.

\item {\bf Javed Miandad} is the greatest batsman Pakistan has ever produced. He played $124$ Test matches for Pakistan, scoring $8832$ runs including $23$ centuries and averaging $52.57$ in Test cricket.

\item {\bf  Imran Khan}  was the finest cricketer Pakistan has produced. He was great all-rounder scoring $3807$ runs as a batsman and claiming $362$ wickets as a bowler in $88$ Test matches he played for Pakistan.

\item {\bf Inzamam ul Haq} has appeared in $120$ Test matches for Pakistan scoring  $8830$ runs which includes $25$ centuries, highest scoring being $329$.

\item {\bf Mushtaq Mohammad} played in $57$ Test matches for Pakistan scoring $3643$ runs with a batting average of $39.17$. His highest score is $201$

\item {\bf Saleem Malik }  is regarded as a gifted cricketer from Pakistan, who played in $103$ Test matches scoring $5768$ runs which includes $10$ centuries. 

\item {\bf Wasim Akram} of Pakistan is rated by many as the best left-arm fast bowler of all time, and his career record certainly speaks volumes about his abilities. In Test cricket he took $414$ wickets at an average of $23.62$, while in the ODIs he took $502$ wickets with an average of $23.52$. 

\item {\bf Hanif Mohammad}  was the first star of Pakistan cricket, who played the longest innings in Test history -  970-minute $337$ runs scored  against West Indies in 1957-58. His versatility extended to captaining and keeping wicket, and bowling right- and left-handed in Test cricket.

\item {\bf GA Headley} was a great batsman and set the standards for generations of West Indian players to follow. He played in $22$ Test matches scoring $2190$ runs with an average of $60.83$. 

\item {\bf LR Gibbs} is regarded as one of the greatest bowlers in history of Test cricket. He played in $79$ Test matches for West Indies dismissing $309$ batsmen. 

\item {\bf CL Walcott} played in $44$ Test matches for West Indies scoring world record of $3798$ runs which includes $15$ centuries. 

\item {\bf S Chanderpaul} is only the second batsman from West Indies who scored more than $10000$ Test runs which includes $27$ centuries, highest score being $203$. 

\item {\bf DL Haynes }  was an opening batsman for West Indies who scored $7487$ runs with a batting average of $42.29$.

\item {\bf CH Lloyd} was a crucial ingredient in the rise of West Indian cricket. He was a hard-hitting left-handed batsman and one of the most successful captains in history of cricket having won two World Cups in $1975$ and $1979$. With an astute tactical brain he led the West Indies to the top of world cricket for two decades. 

\item {\bf Sir GS Sobers} of West Indies scored $8032$ in Test cricket which includes $22$ centuries. He averages $57.78$ in Test cricket. Apart from being an excellent batsman and a great left-arm slow bowler, he was an enterprising captain. 

\item {\bf IVA Richards} of West Indies scored $8540$ runs in Test cricket, including $24$ centuries at an average of $50.23$. The most feared batsman of his era, his power was awesome,  and never wore a helmet.
 
\item {\bf FMM Worrell}  was West Indies'  first appointed black captain and was also their most charismatic and influential. Though a fine, stylish batsman having scored $3860$ runs in $51$ Test matches, it is as a strong captain and an uniting force that he is  remembered.  
\end{itemize}

\end{document}